\documentclass[12pt, a4paper]{article}
\usepackage{graphicx}

\begin{document}

\title{Efficient quantum computation\\ within a disordered Heisenberg
spin-chain}
\author{Chiu Fan Lee\thanks{c.lee1@physics.ox.ac.uk}
\ and \ Neil F. Johnson\thanks{n.johnson@physics.ox.ac.uk}
\\
\\ Centre for Quantum Computation and Physics Department \\ Clarendon
Laboratory, Oxford University \\ Parks Road, Oxford OX1 3PU, U.K.}

\maketitle

\abstract {We show that efficient quantum computation is possible using a
disordered Heisenberg spin-chain with `always-on' couplings. Such disorder
occurs naturally in nanofabricated systems. Considering a simple chain setup,
we show that an arbitrary two-qubit gate can be implemented using just three
relaxations of a controlled qubit, which amounts to switching the on-site
energy terms at most twenty-one times. }

\newpage

Many possible physical implementations for performing quantum  computation have
been proposed (see Ch.8 in \cite{NC00} for example).  However the vast majority
of these schemes require that the  coupling between qubits (e.g. two-level atoms
or quantum dots) can be turned on  and off. In addition to the difficulty of
doing this sufficiently quickly so as to avoid decoherence, and sufficiently
accurately so as not to introduce  unwanted errors, there is the fundamental
problem that coupling terms in the Hamiltonian  of multi-qubit systems are
permanently `on', i.e. they are  time-independent.  Building upon
the work of Refs.
\cite{DBK00,ZZG02}, Benjamin and Bose \cite{BB03} recently made a  significant
advance beyond this paradigm by introducing a simple method for performing 
quantum computation with a Heisenberg spin-chain where the spin-spin couplings
are `always on', i.e. constant in time.  This scheme is clearly of great
potential use -- however, a quantitative study of its efficiency and possible
generalization has not yet been reported. Futhermore the couplings between
each spin-spin pair were assumed to be identical \cite{BB03,Ben04,BB03b} --
however, this situation cannot be engineered reliably in systems such as
nanostructure arrays.

In this paper, we generalize Benjamin and Bose's scheme to a Heisenberg spin
chain with non-identical couplings between spins (i.e. qubits). Our scheme of
time-independent, non-identical qubit-qubit couplings, will arise naturally  in
nanofabricated systems such as arrays of quantum dots, or chains of C$_{\rm
60}$ buckyball cages. It can also be engineered to arise for atoms in optical
traps.  Within a simple chain setup, we show that an arbitrary two-qubit gate
can be implemented reliably with at most three relaxations of one controlled
qubit. Such two-qubit gates are a crucial ingredient for universal quantum
computation. This number of relaxations amounts to switching the on-site
energies at most twenty-one times. The major difficulty within this type of
quantum computing scheme arises when the controlled qubits are relaxed.
However, our work shows that the number of times the controlled qubits are
relaxed can be minimized using a simple searching procedure.  Our findings
should provide  considerable comfort for experimentalists preparing to build
candidate quantum computer systems using coupled nanostructures: Time, effort
and resources do not have to be wasted on engineering identical qubit-qubit
couplings.

The Hamiltonian we consider is the following:
\begin{equation} 
\label{define_H} H=\sum_j E_j(t) \sigma^Z_j + \sum_{j} J_{j,j+1}
\underline{\sigma}_j \cdot \underline{\sigma}_{j+1}
\end{equation}  where $j$ labels the qubit (e.g. quantum dot, C$_{\rm 60}$
buckyball or atom) and the qubit-qubit couplings are represented by
$\{J_{j,j+1}\}$. For simplicity we consider the case where the $J$'s in each
pair
$(J_{j-1,j}, J_{j,j+1})$ do not differ by more than ten percent in magnitude --
however there is no indication that similar results will not hold outside this
range.
Recall that single-qubit operations can be implemented if the
following sum of terms
\begin{equation}
\label{additional}
\sum_j \left[ F_j(t) \sigma^X_j + G_j(t) \sigma^Y_j \right]
\end{equation} is added to the above Hamiltonian. It is assumed that
the experimentalist has control over the temporal form of the
on-site energies
$\{E_j(t), F_j(t), G_j(t) \}$. For example, these terms can be controlled by
applying a magnetic field (hence changing the Zeeman energy) or electric field
(hence introducing a Stark shift) to the qubit $j$ \cite{LD98} in the 
appropriate direction. We will henceforth assume that single-qubit operations
are possible (which is essentially the  same assumption as adopted in Refs. 
\cite{ZZG02, BB03}) and will concentrate on two-qubit operations. Hence we will
discard the single qubit terms given in  Equation \ref{additional},
since they are not needed for the discussion of two-qubit operations.

Consider three arbitrary qubits within this chain. We assume that qubits 1 and
3 are work qubits and that  the qubits 0 and 4 adjacent to them are fixed. 
Without loss of generality, we can incorporate these interactions into the
on-site energy terms of  qubits 1 and 3. Focusing on qubits 1 to 3, we note
that the four subspaces spanned by $\{|000
\rangle\}$,
$\{|001
\rangle, |010 \rangle, |100 \rangle\}$, $\{|110 \rangle, |101
\rangle, |011 \rangle\}$ and $\{|111 \rangle \}$ are invariant under $H$.
Within the basis
$\{|001 \rangle,|010
\rangle, |100 \rangle\}$, the operation $H$ is:
\begin{equation} U \equiv
\left( \begin{array}{ccc} A_1 & 2J_{23} & 0 \\ 2J_{23} & A_2 & 2J_{12} \\ 0 &
2J_{12} & A_3
\end{array} \right)
\end{equation} where
\begin{eqnarray} A_1 &=& E_1+E_2-E_3 + J_{12}-J_{23} \\ A_2 &=& E_1-E_2+E_3 -
J_{12}-J_{23}
\\ A_3 &=& -E_1+E_2+E_3-J_{12}+J_{23}\ \ \ .
\end{eqnarray} Within the basis $\{|110 \rangle,|101 \rangle, |011
\rangle\}$, the operation $H$ is:
\begin{equation} V \equiv
\left( \begin{array}{ccc} B_1 & 2J_{23} & 0 \\ 2J_{23} & B_2 & 2J_{12} \\ 0 &
2J_{12} & B_3
\end{array} \right)
\end{equation} where
\begin{eqnarray} B_1 &=& -E_1-E_2+E_3 + J_{12}-J_{23} \\ B_2 &=& -E_1+E_2-E_3 -
J_{12}-J_{23}
\\ B_3 &=& E_1-E_2-E_3-J_{12}+J_{23}\ \ \ .
\end{eqnarray} We assume that qubit 2 is initialized to $|0 \rangle$. To
perform a two-qubit operation on qubits 1 and 3, qubit 2 has to be decoupled
from them after the operation. Namely, $[e^{-iU\tau}]_{12}, [e^{-iU\tau}]_{21},
[e^{-iU\tau}]_{23}$ and $[e^{-iU\tau}]_{31}$ are all zero, where $\tau$ is the
duration of the operation and $[e^{-iU\tau}]_{ij}$ denotes the $(i,j)$-entry of
$e^{-iU\tau}$ in matrix form. The same requirements apply to the matrix 
$[e^{-iV\tau}]$. 
Due to  unitarity of the operation, this amounts to requiring that the
magnitudes of
$[e^{-iU\tau}]_{22}$ and $[e^{-iV\tau}]_{22}$ equal one. If this is achieved,
then the corresponding two-qubit operation will be:
\begin{equation}
\label{main_matrix} N \equiv
\left( \begin{array}{cccc}
\alpha & 0&0&0 \\ 0& [e^{-iU\tau}]_{11} & [e^{-iU\tau}]_{13} & 0 \\ 0&
[e^{-iU\tau}]_{31}& [e^{-iU\tau}]_{33} & 0 \\ 0 & 0 & 0&\beta
\end{array} \right)
\end{equation} where
\begin{eqnarray}
\alpha &=& e^{-i(E_1+E_2-E_3 + J_{12}+J_{23}) \tau} \\
\beta &=& [e^{-iV\tau}]_{22} \ .
\end{eqnarray}

\begin{figure}

\caption{(color online) Results obtained using the algorithm in the main text 
with
$J_{12}=1$, $L=5$ and $m=30$.
$\triangle$ corresponds to the mean separation between two neighboring  data
points in terms of their classifying angles (left axis), $\bigcirc$ corresponds
to largest separation between two neighboring  data points in terms of their
classifying angles (left axis); $\times$ corresponds to the number of data
points  collected (right axis).}
\label{heisenberg_1}
\begin{center}
\includegraphics[scale=0.65]{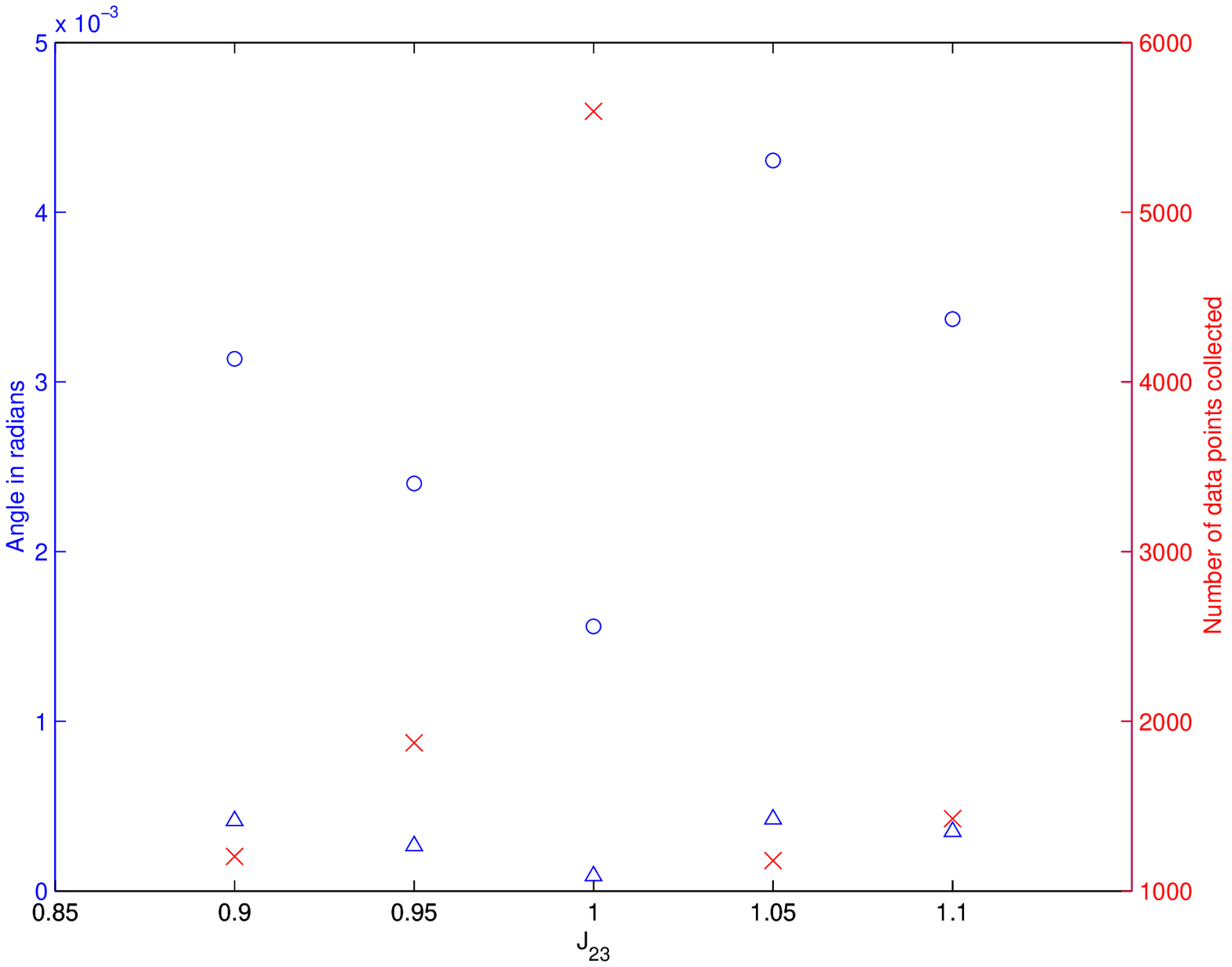}
\end{center}

\end{figure}

Our aim now is to deduce a set of $(\{E_j\}, \tau)$ values which will allow us
to decouple the controlled qubit after time evolution $\tau$. Namely, we are
searching for solutions that satisfy the two constraints:
$[e^{-iU\tau}]_{22}=1$ and $[e^{-iV\tau}]_{22}=1$, in a four dimensional space.
The use of various optimization techniques in gate-building in the classical
domain has been explored in Ref. \cite{WM04}. Here, we employ the gradient
descent method with random initial positions, which is common to all control
optimization problems (see Ch. 9 in Ref.
\cite{BV04} for example). Our search algorithm goes as follows:
\begin{enumerate}
\item Given a range $L$ and density parameter $m$, we divide the set $\{E_j\}$
into $m$ values $\{-L+ \frac{2Lk}{m-1} : 0\leq k\leq m-1\}$ and we divide
$\tau$ into $m$ values $\{ L+\frac{3Lk}{m-1} : 0\leq k\leq m-1\}$.
\item For each of the $m^4$ point combinations, we calculate the corresponding
matrices $U$ and
$V$. These constitute the initial position of our gradient descent method to
find the parameters $(\{E_j\}, \tau)$, with objective function:
\begin{equation} {\rm Obj}(\{E_j\}, \tau)  \equiv
\left|[e^{-iU\tau}]_{22} \right|^2 +\left| [e^{-iV\tau}]_{22} \right|^2.
\end{equation} We then proceed with the gradient descent method. Specifically,
we first fix
$\tau$ and perform a descent with respect to $\{E_j\}$, then we fix the set
$\{E_j\}$ and perform a descent on the coordinate
$\tau$. We repeat the process until a local minimum is attained --
specifically, we repeat the process until the change in the ojective function is
less than 0.000005. If the value of the objective function at that local
minimum is higher than
$1.99995$, we form the matrix $N$ as defined above and calculate its
classifying angles as in Refs. \cite{KC01,ZVS03a}.
\item If the angle is of the form $(0,0,c)$ with uncertainty $\pm 0.0025 \pi$,
it is recorded.
\end{enumerate} Step 1 partitions the search problem into many possible
starting positions, in order that different two-qubit operations can be
generated. The search in Step 2 ensures that the combination $(\{E_j\}, \tau)$
decouples the controlled qubit from the work qubits as much as possible. We
note that we separate the descents with respect to the set $\{E_j \}$ and to
the coordinate $\tau$ because their effects on the objective function are very
different. The set $\{E_j\}$ affects the eigenvalues and eigenvectors of
the matrices $U$ and
$V$ while 
$\tau$ acts as a scalar multiplier of the eigenvalues.  The results are shown in
Figure~\ref{heisenberg_1} with $L=5$ and $m=30$. For the coupling constants
$(J_{12},J_{23})
\in \{ (1,0.9),(1,0.95),(1,1.05),(1,1.1)\}$, we find that {\em almost all}
values of the rotation angle in the interval $[0,\pi/2]$ become represented.
The largest gap between two nearest-neighbor points in terms of the rotation
angle is
$\sim 0.0045\pi$, while the mean gap between two nearest neighboring points in
terms of the rotation angle is less than
$0.0005\pi$. It takes about 8 hours of computing time for each $J$
value on a standard laptop computer. Since the algorithm is highly
parallel, the computing time will be cut by one half if two computers are on
and so forth. The particular coupling constants  shown in
Figure
\ref{heisenberg_1} are chosen to illustrate the algorithm -- further
simulations indicate that similar results should hold for other sets of
values.  In short, our simulations demonstrate that by controlling the
on-site energies in a Heisenberg spin-chain, one can generate two-qubit gates
of the form $e^{ci\sigma_z^1 \sigma_z^2}$ with arbitrary
$c$ with some tolerable errors.  Specifically, our simulations suggest the
following conjecture:
\begin{quote} {\bf Conjecture.}  For all non-zero $J_{12}$ and $J_{23}$, for any
$c \in [0,\pi/2]$ and for all $\epsilon > 0$, there exists a set
$(\{E_j\}, \tau)$ such that 
\begin{equation}
\left| {\rm Tr}_2 e^{-iH(\{E_j\})\tau} - e^{ci\sigma^Z \otimes \sigma^Z}
\right| < \epsilon
\end{equation} where Tr$_2(.)$ denotes the operation of tracing out the second
qubit,
$H(\{E_j\})$ is as shown in Equation \ref{define_H}, and $|M|$ returns the
absolute value of the entry with the maximum norm in the matrix $M$.
\end{quote} We have not been able to formally prove or disprove this conjecture.
However we recall that our algorithm  corresponds to searching for
four-dimensional vectors $(\{E_j\}, \tau)$ which satisfy two constraints:
$[e^{-iU\tau}]_{22}=[e^{-iV\tau}]_{22}=1$. Hence we speculate that this excess
in degrees of freedom  will enable us  to perform different two-qubit
operations. 

References \cite{KC01,ZVS03a,ZVS03b} showed that an arbitrary entangling
gate $Q$ can be written as $Q=S_l A S_r$, where
\begin{equation} A=e^{c_1i\sigma^X_1 \sigma^X_2}\ e^{c_2i\sigma^Y_1
\sigma^Y_2} \ e^{c_3i\sigma^Z_1 \sigma^Z_2}
\end{equation} with $S_l,S_r \in {\rm SU(2)} \otimes {\rm SU(2)}$. Since
\begin{equation} R_k e^{ci\sigma^k_1 \sigma^k_2} R_k^\dag = e^{ci\sigma^Z_1
\sigma^Z_2}
\end{equation} where $R_k = e^{(\pi/4)i \sigma^k}\otimes e^{(\pi/4)i \sigma^k} 
\in {\rm SU(2)} \otimes {\rm SU(2)}$ and 
$k \in \{ X, Y \}$, we see that an arbitrary two-qubit gate can be implemented
by letting the controlled qubits relax three times at most. This amounts to
switching the on-site energy terms twenty-one times at most.  For example,
since the CNOT can be represented as
$(0,0,\pi/4)$ \cite{ZVS03a}, it can be implemented in the present scheme with
just one relaxation of the controlled qubit. We note that our measure of
efficiency relates directly to the difficulties arising in  physical
implementations, i.e. the difficulties in relaxing the controlled qubits.
Hence, the  fewer number of times the controlled qubit need to be relaxed, the
more efficient the computing  scheme is. On the other hand, since non-trivial
entangling only occurs when the controlled qubit  is relaxed, one can easily
relate our measure of efficiency to the number of entangling operations 
needed, which is the usual measure of efficiency in quantum  computing (see
Refs. \cite{ZVS03b, VD04} for example). In this regard, our method compares
well with the  most recent result obtained in Ref. \cite{VD04} since our
results also indicate an upper bound of three  relaxations/entangling
operations to implement an arbitrary two-qubit gate.

\begin{figure}
\caption{(color online) Profile showing the switching of $\{E_i\}$ as a
function of time, in order to realize a generic two-qubit gate. 
Red dotted line is $E_1$; green broken line is $E_2$; blue
solid line is $E_3$ The sequence
shown requires a total of twenty-one switchings of the on-site energies. An
energy value represented by the word `Passive' is assumed to be sufficiently
large such that qubit 2 is fixed at state $| 0 \rangle$.  }
\label{heisenberg_3}
\begin{center}
\includegraphics[scale=.7]{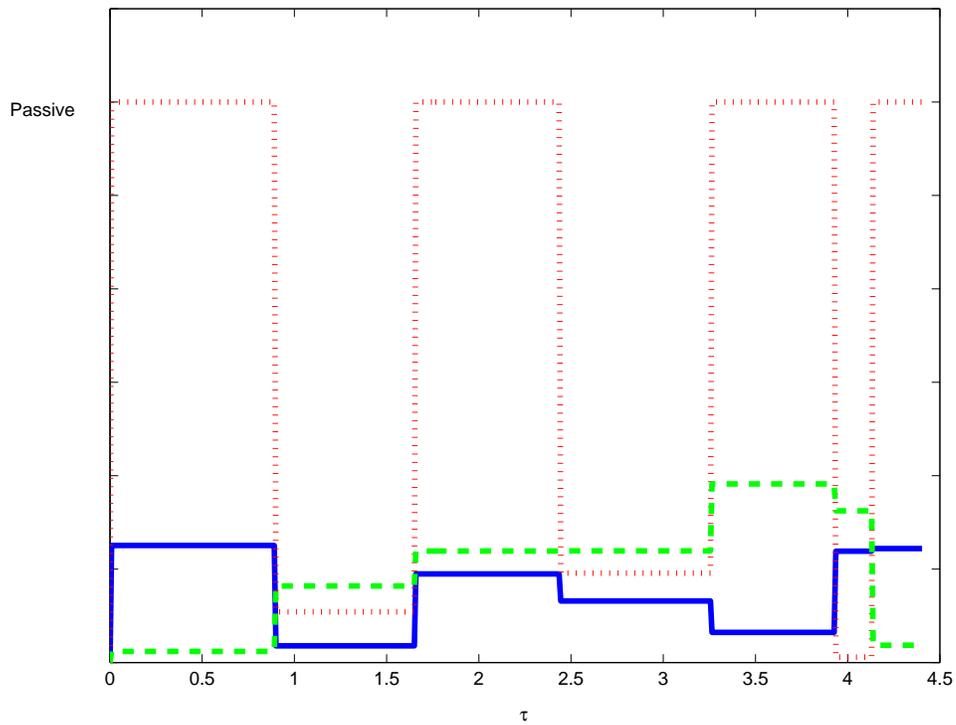}
\end{center}
\end{figure}

\begin{figure}

\caption{Plot of a typical evolution of the objective function versus the
number of descents performed. Final value of the objective function in this
simulation is greater than $1.999999$  and the classifying angles produced are
$(0,0,0.4159\pi)$.}
\label{heisenberg_2}
\begin{center}
\includegraphics[scale=0.7]{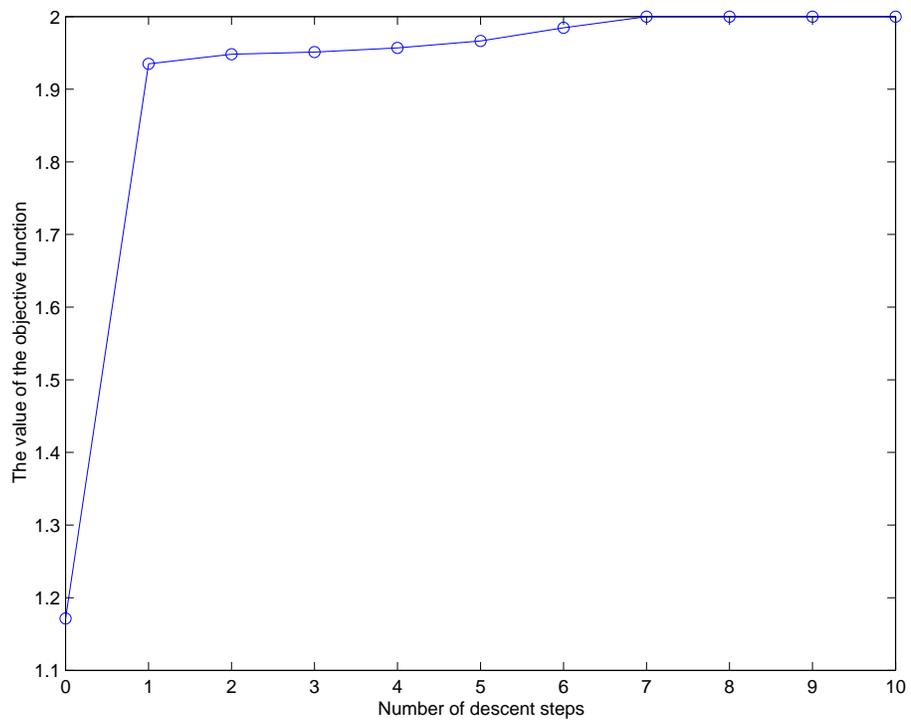}
\end{center}
\end{figure}

Figure \ref{heisenberg_3} shows a systematic plot of the switching profiles
which illustrate this result. We note that the error in decoupling qubit 2
from qubit 1 and 3 could be further reduced by running longer simulations, as
shown in Figure \ref{heisenberg_2}.   The error in coverage of the angle
range can also be improved with more simulation. Its presence can also be seen
as making the quantum computation more probabilistic and hence should not be
viewed as being detrimental.

In conclusion, we have provided a generalization of the quantum computing
scheme introduced in Ref. \cite{BB03} and have also discussed its efficiency.
We have found the important practical and theoretical result that identical
qubit-qubit couplings are {\em not} necessary for universal quantum
computation, thereby eliminating a fierce requirement on the precision of
nanofabrication. We believe that the results of this work are relevant to a
range of quantum computing implementations, and should stimulate
experimentalists to explore less mature fabrication technologies. In
particular, the  flexibility of the objective function in the gradient descent
step should allow experimentalists to weigh parameters differently in order to
reflect particular experimental conditions, or characteristics of fabrication.
For example if the length of time for temporal evolution is a major constraint,
perhaps because of decoherence processes, a penalty term of the form
$-\tau$ could be added to the objective function in order to skew the
optimization.

In more general terms, the method introduced here is not restricted to the
Heisenberg  interaction Hamiltonian. Specifically,  given a set of fixed
parameters and a set of controllable parameters which would be  determined by
physical requirements, one could search over the controllable parameters  to
try to come up with specific gates, as in the present paper. Therefore we
envisage usefulness for this scheme when applied to more general Hamiltonian
systems.

\indent C.F.L. thanks University College (Oxford) and NSERC
(Canada) for financial support. N.F.J. thanks the LINK-DTI (UK)
project. The authors are also very grateful to Simon Benjamin for useful
comments.

\newpage

\end{document}